\documentstyle[aps,prl,preprint]{revtex}
\tighten
\begin{document}
\preprint{UTPT-93-32, UM-P-94/18}
\draft
\title{Fractal basins and chaotic trajectories in multi-black hole space-times}
\author{C. P. Dettmann and N. E. Frankel}
\address{School of Physics, University of Melbourne,
 Parkville Victoria 3052, Australia}
\author{N. J. Cornish}
\address{Relativity Group, Department of Physics, University of Toronto,\\
Toronto, Ontario M5S1A7, Canada}
\maketitle
\begin{abstract}
We investigate the phase-space for trajectories in multi-black hole
spacetimes. We find that complete, chaotic geodesics are well described by
Lyapunov exponents, and that the attractor basin boundary scales as a
fractal in a diffeomorphism invariant manner.
\end{abstract}
\narrowtext

Chaos is an aspect of General Relativity which has only recently been explored.
Its existence is expected as General Relativity contains
generalizations of well-known chaotic systems from Newtonian physics such
as the 3-body problem. Moreover, the non-linearity of Einstein's
equations may give rise to chaos in systems whose Newtonian analogue
is non-chaotic. The characterization of chaos in General Relativity
is complicated by the dynamical nature of time, which necessitates a careful
generalization of the parameters which quantify chaos, such as Lyapunov
exponents.

Relativistic systems in which chaos is known
include charged particles in a magnetic field interacting with gravitational
waves~\cite{VP} and particles near a magnetized black hole~\cite{KV}, as well
as particles in Majumdar-Papapetrou (MP) geometries, described below. There
is also the general, sufficient but not necessary, criterion that chaotic
geodesics will occur if the phase space is compact and the Ricci scalar is
negative~\cite{SSB}. In MP spacetimes the Ricci scalar is identically zero,
so this criterion is inapplicable.

Chaos is also found in cosmological models such as the Mixmaster universe%
~\cite{Mis},~\cite{Z},~\cite{H} and Robertson-Walker models containing
dynamical fields~\cite{CH},
however there is difficulty in defining coordinate invariant measures of
chaos for such systems~\cite{BT}.

MP geometries~\cite{M},~\cite{P} are the static solution of the
Einstein-Maxwell equations with metric and electrostatic potential given by
\begin{eqnarray}
ds^{2}=-U^{-2}dt^{2}+U^{2}(dx^{2}+dy^{2}+dz^{2})\;\;,\\
A_{t}=U^{-1}\;\;,
\end{eqnarray}
and the spatial components of $A_{\mu}$ are zero, with $U$ a function
of the spatial coordinates satisfying Laplace's equation
\begin{equation}
\nabla^{2}U(x,y,z)\equiv U_{,xx}+U_{,yy}+U_{,zz}=0\;\;.
\end{equation}
Hartle and Hawking~\cite{HH} showed that if $U$ is of the form
\begin{equation}
U=1+\sum_{i=1}^{N}\frac{M_{i}}
{\sqrt{(x-x_{i})^2+(y-y_{i})^2+(z-z_{i})^2}}\;\;,
\end{equation}
the MP metric corresponds to a system of extreme Reissner-Nordstrom black holes
with equal charge and mass $M_{i}>0$ and horizons at $(x_{i},y_{i},z_{i})$.
Note that the MP coordinates are singular at these points, mapping a horizon
of finite proper area to a single point. They also showed that if $U$ takes
a form different to the above expression, the space-time contains naked
singularities.

Chandrasekhar~\cite{Candra} and Contopoulos~\cite{C90},~\cite{C91} have
investigated the timelike and null
geodesics of the two black hole system from the point of view of the periodic
orbits and the weak field limit. For particles with elliptic (bound)
energies the trajectories fall into several categories. There are stable
periodic and quasiperiodic orbits, chaotic orbits trapped between periodic
orbits, trajectories which fall into one or other of the black holes, and
chaotic trajectories which lie on the boundary of these regions. This
structure, together with the fact that the weak field limit is integrable,
makes the MP geodesics problem a particularly interesting example of chaos in
General Relativity.

Working from the equations of motion, Contopoulos showed
that the weak field limit of the two black hole system was integrable
for uncharged test particles with zero angular momentum about the
axis of symmetry. Using Hamilton-Jacobi methods~\cite{Carter},
we have generalized this result to include test particles of arbitrary
charge, energy and angular momentum. In the background spacetime
generated by two black holes with masses $M_{1}$, $M_{2}$ centered at
$(0,0,0)$ and $(0,\pi,0)$ in prolate spheroidal coordinates,
$(\psi, \theta, \phi)$~\cite{C90}, we find the Super-Hamiltonian for a
test particle of charge $e$ and mass $m$ is given by
\begin{equation}
{\cal H}={\partial S \over \partial \lambda}={1 \over 2}U^{2}\left({\partial
S \over \partial t}-{e \over U}\right)^{2}
-{1\over 2U^{2}Q}\left[
\left({\partial S \over \partial \psi}\right)^{2}
+\left({\partial S \over \partial \theta}\right)^{2}\right]
-{1 \over 2(U\sinh\psi \sin\theta)^{2}}\left(
{\partial S \over \partial \phi}\right)^{2}
 \; , \label{HJ}
\end{equation}
where
\begin{eqnarray}
U&=&1+W/ Q \; , \hspace{0.2in} 
Q=\sinh^{2}\psi +\sin^{2}\theta \; , \nonumber \\
W&=&(M_{1}+M_{2})\cosh\psi+(M_{1}-M_{2})\cos\theta \; . \nonumber
\end{eqnarray}
Since $t$ and $\phi$ are cyclic coordinates, the canonically conjugate
momenta $\pi_{t}=-(E-e)$, and $\pi_{\phi}=L_{z}$ are constants of the
motion. $E$ and $L_{z}$ are the particle's energy and angular momentum at
infinity. Invariance of the particle's rest mass along the trajectory affinely
parameterized by $\lambda$ gives rise to a third constant of motion, $m^2$.
We see that the system is separable, and hence integrable, if
\begin{equation}
S=-{1 \over 2}m^2\lambda-(E-e)t+L_{z}\, \phi+S_{\psi}(\psi)
+S_{\theta}(\theta) \; .
\end{equation}
Substituting this expression into (\ref{HJ}) and using the 
weak field expansion
\begin{equation}
QU^{n}=Q+nW+{\cal O}(W^{2}/Q) \; ,
\end{equation}
we find
\begin{eqnarray}
S_{\psi}&=&\int \left((E^2+m^2)\sinh^{2}\psi
+2(M_{1}+M_{2}) (2E^2+m^2-Ee)\cosh\psi -{L_{z}^2
\over \sinh^{2}\psi}+\alpha \right)^{1/2} d\psi \; , \\
S_{\theta}&=&\int \left((E^2+m^2)\sin^{2}\theta
+2(M_{1}-M_{2}) (2E^2+m^2-Ee)\cos\theta -{L_{z}^2
\over \sin^{2}\theta}-\alpha \right)^{1/2} d\theta \; .
\end{eqnarray}
The separation constant $\alpha$ is a fourth constant of the motion,
which in turn guarantees the system is integrable.
 
The next order in the weak field expansion destroys separability in prolate
spheroidal coordinates. While this does not prove the system is non-integrable,
it does show that if chaos is present, it is due to relativistic effects.
In this case, the chaos is related to the relativistic perihelion advance of
elliptic trajectories, which invalidates Bonnet's theorem
for motion with two centres \cite{Candra}.

For MP geometries with three or more black holes, even the
Newtonian limit is non-integrable. The one exception to this occurs in the
Newtonian regime for ``extremal'' test particles with $e=m$, since then the
gravitational and electrostatic forces cancel exactly. This cancellation
breaks down in General Relativity where gravity couples to all forms
of energy, including kinetic energy. The scattering of an extremal test
particle by an extremal black hole of mass $M$ illustrates this effect
since we find the scattering angle $\Delta \phi$ differs from the
Newtonian value ($\Delta \phi =0$) by
$\Delta \phi = -(M / b)[3+u^2 /4+ {\cal O}(u^4 , M/b)]$.
The impact parameter $b$ is related to $L_{z}$ and the asymptotic
velocity $u$ via $L_{z}= mbu/ \sqrt{1-u^2}$. This result leads us to
expect that the trajectories of extremal test particles in
MP geometries will be chaotic beyond the Newtonian limit, thus
providing another example where chaos arises due to relativistic effects.

For rapid numerical calculations we use the original MP coordinates
$(t,x,y,z)$, because the equations of motion do not contain transcendental
functions.
Writing the upper components of the four-velocity in an orthonormal
basis as $(v_{0},{\bf v})$, the equations of motion are
\begin{eqnarray}
\dot{\bf v}&=&U^{-2}[(v_{0}^{2}+v^{2}-ev_{0}/m)\nabla U-
{\bf vv}\cdot\nabla U]\;\;,\\
\dot{\bf x}&=&U^{-1}{\bf v}\;\;,\\
\dot{t}&=&Uv_{0}\;\;,\\
v_{0}&=&\sqrt{1+v^{2}}\;\;,
\end{eqnarray}
where a dot indicates the derivative with respect to proper time.
The conserved energy is given by
\begin{equation}
E=U^{-1}(mv_{0}-e)\;\;.
\end{equation}
\includegraphics{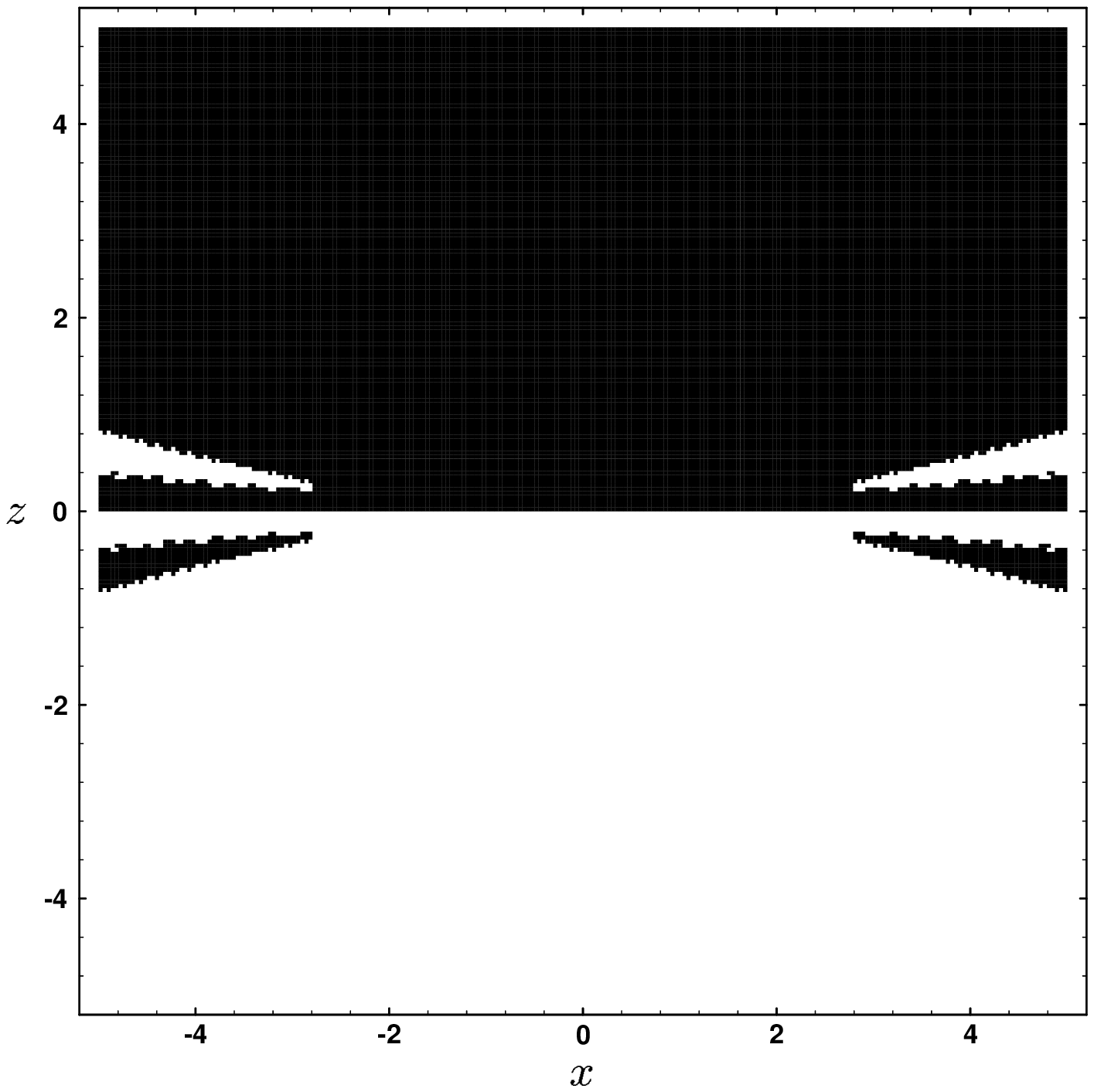}
\vspace{90mm}

FIG. 1.
{\small A two-dimensional section of phase space determined by $\bf v=0$,
$y=0$.  The black regions are trajectories which fall into the black hole of
mass $1/3$ at $(0,1)$.  The white regions are those which fall into the black
hole of mass $1/3$ at $(0,-1)$.}

\vspace{5mm}
We have integrated the equations using a 4th order Runge-Kutta technique with
adaptive stepsize~\cite{PFTV}. The results are shown in Figs.1--3.
These figures show two-dimensional sections of the phase space given by the
initial conditions ${\bf v}={\bf 0}$, $y=0$.
All of the points shown fall into one of the black
holes and are color coded accordingly. The boundary between the basins of
attraction clearly has a complicated structure, which may be quantified
using the concept of fractal dimension.
\vspace{100mm}

\includegraphics{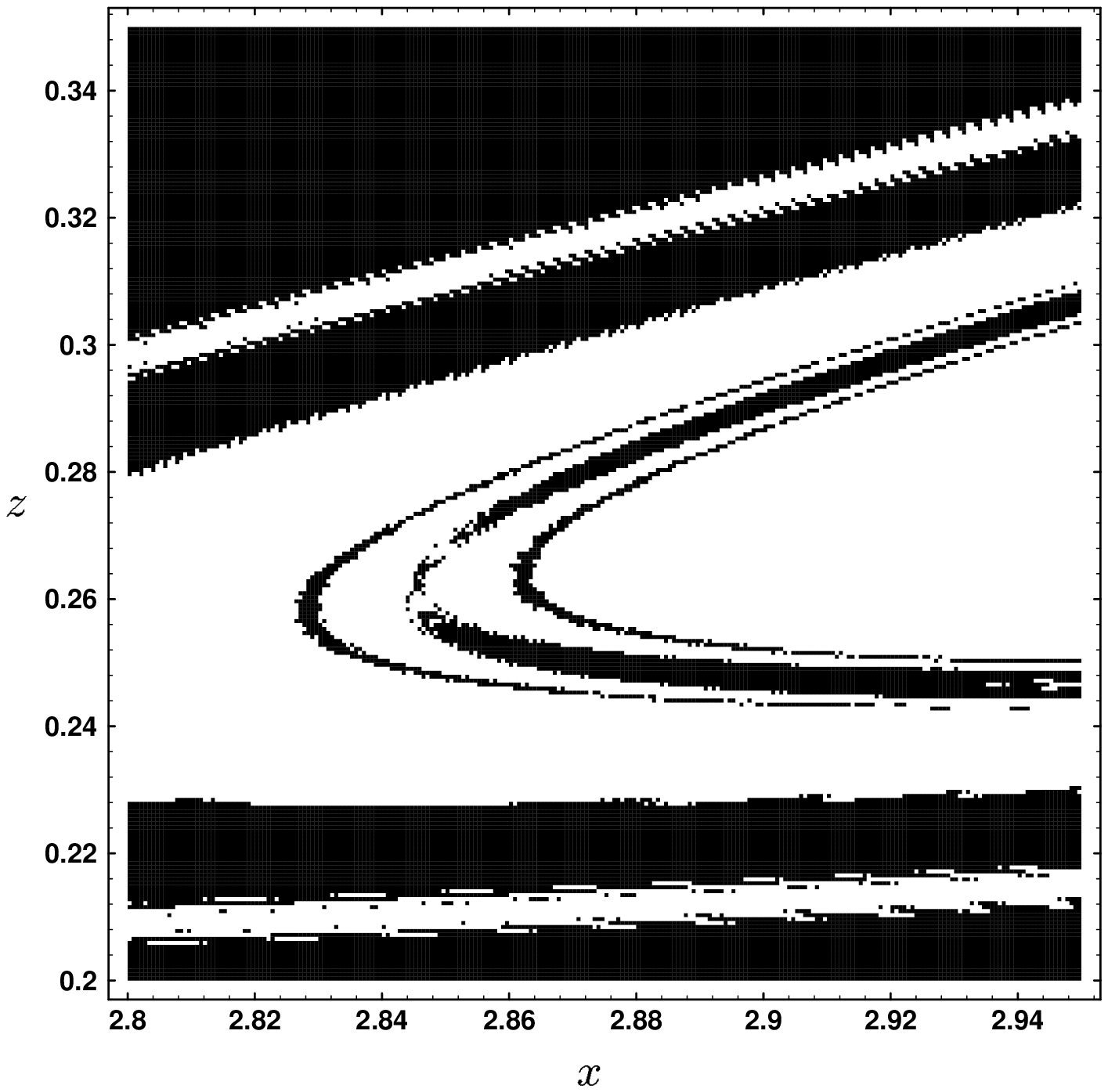}
\centerline{ FIG. 2. {\small A small region of Fig. 1 showing the fractal
structure.}}
\vspace{5mm}

There are a number of different fractal dimensions defined in the
literature~\cite{F}. The box-counting dimension of a set embedded in an
$E$-dimensional Euclidean space is obtained by finding the minimum number
$N(\epsilon)$ of $E$-dimensional cubes needed to cover the set, and
calculating 
\begin{equation}
d_{B}=-\lim_{\epsilon\rightarrow0}\frac{\ln N}{\ln\epsilon}\;\;,
\end{equation}
assuming that this limit exists. The generalization to Riemannian or
pseudo-Riemannian spaces is direct, since $d_{B}$ is invariant under
diffeomorphisms, so all choices of coordinates give the same value.
If the basin is defined on some spatial hypersurface, as is the case
here, the dimension may depend on the chosen slicing of spacetime.

The value of $d_{B}$ of the fractal boundary for the region of phase space
shown in Fig. 2 was evaluated using the above equation. The region,
containing $2520^2$ points, was covered by a grid. Each square, of size
$\epsilon^2$ ($\epsilon$ being a factor of $2520$) was counted if it contained
trajectories leading to both black holes. Fig. 4 shows a plot of $\ln N$ vs
$\ln\epsilon$. The straight line is a fit to all but the three
largest and smallest values of $\epsilon$, and gives a dimension
of $1.464\ldots$.
This result clearly demonstrates that the 8-dimensional boundary as a whole
is a fractal, thus giving us a diffeomorphism invariant measure of chaos. 

\newpage
\vspace*{100mm}
\includegraphics{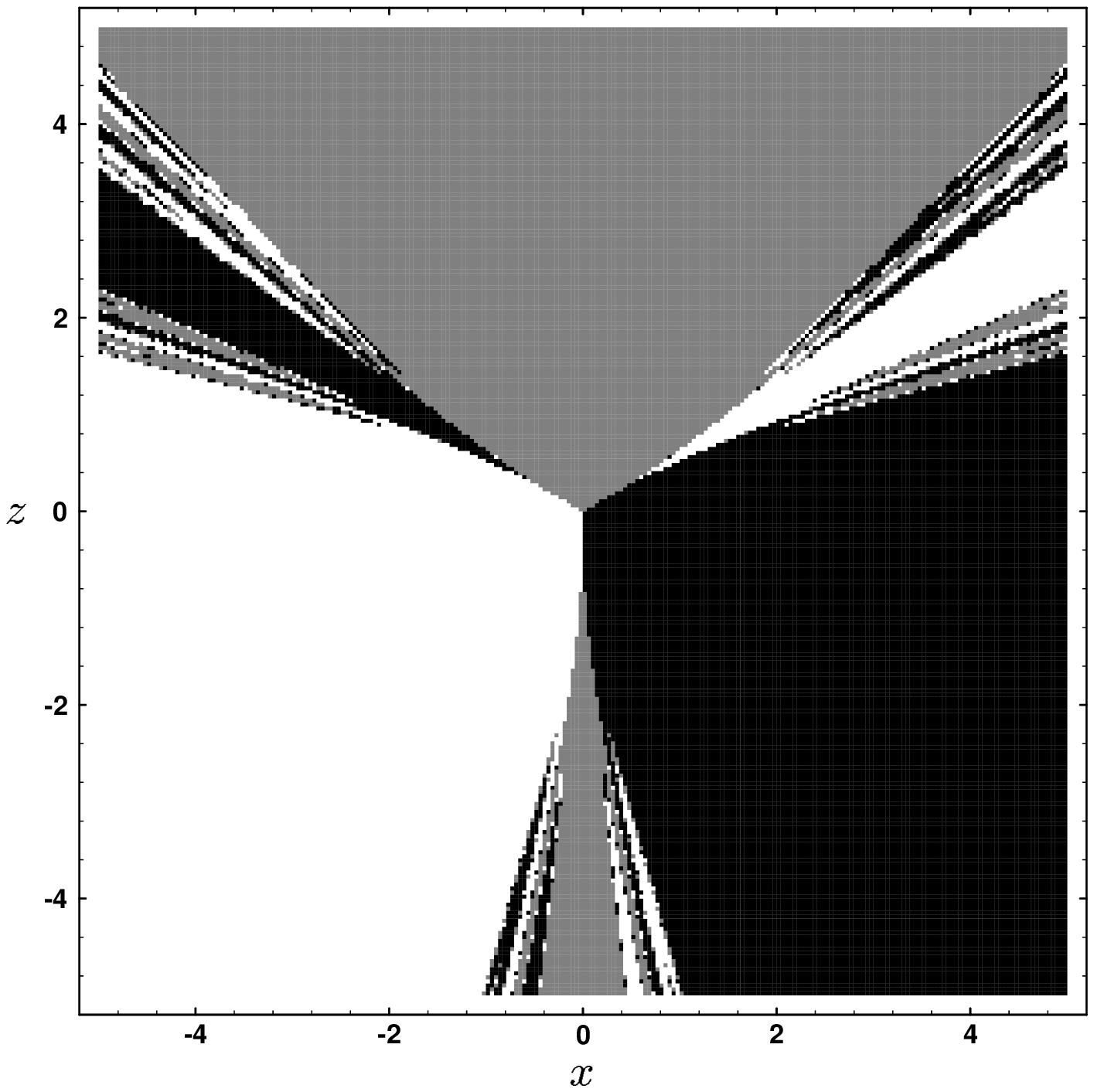}

FIG. 3. {\small The $\bf v=0$, $y=0$ section of phase space for the
3-hole problem. The black holes of mass $1/3$ are situated at $(0,1)$,
$(\protect\sqrt{3}/2,-1/2)$, and $(-\protect\sqrt{3}/2,-1/2)$, corresponding
to gray, black, and white, respectively.}

\vspace*{80mm}

\includegraphics{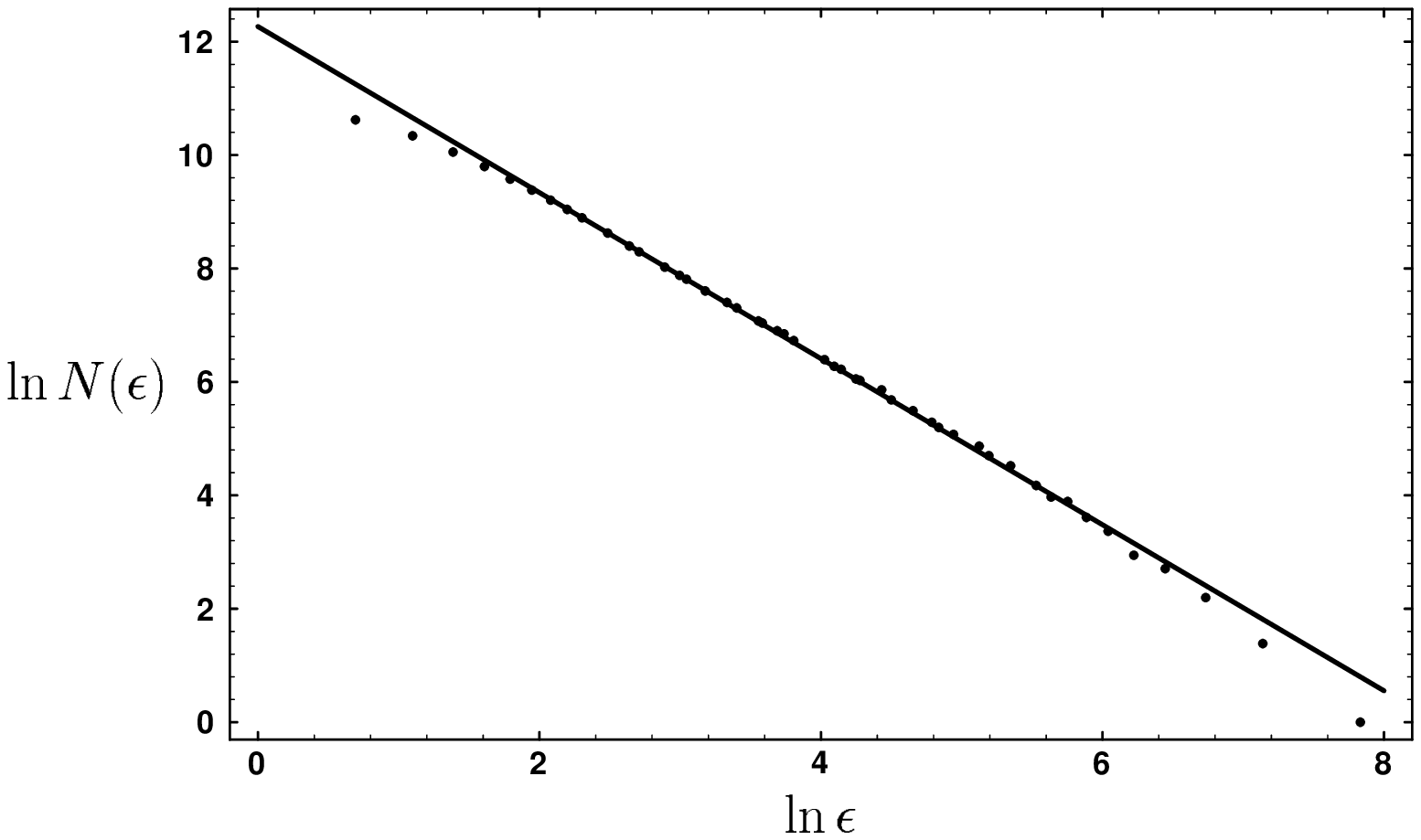}

FIG. 4. {\small The dimension $d_{B}$ of a region of the fractal
boundary near that shown in Fig. 2 is equal to $1.464\ldots$.}
\vspace*{5mm}

For complete geodesics, a useful characteristic of chaotic systems is the
presence of positive Lyapunov exponents, which demonstrate that the system is
nonintegrable~\cite{CCF}. These are defined in flat space-time by choosing a
point $x$ in phase space, at the centre of a ball of radius $\epsilon\ll 1$.
After a time $t$ the ball evolves to an ellipsoid with semi-axes
$\epsilon_{k}(t)$, where $k$ ranges from one to the dimension of the phase
space. The Lyapunov exponents are
\begin{equation}
\lambda_{k}(x)=\lim_{t\rightarrow\infty}\lim_{\epsilon\rightarrow 0}
\frac{1}{t}\ln\frac{\epsilon_{k}(t)}{\epsilon}
\;\;,\label{Liap}\end{equation}
again assuming the limits exist. The $\lambda_{k}$ are constant along a
trajectory, and are often constant over larger regions of phase space, such
as the basin of an attractor. Numerically these are calculated by integrating
the linearized equations and periodically performing a Gram-Schmidt
orthonormalization to ensure that the exponentially growing solutions do not
cause overflow errors~\cite{SN}.

There are a number of subtleties associated with the definition and calculation
of Lyapunov exponents in curved spaces, and in particular, in the MP 
space-times. The definition of the $\lambda_{k}$ requires a metric on phase
space to calculate distances, but the $\lambda_{k}$ are independent of this
metric if the phase space is compact, since the trajectory passes arbitrarily
close to at least one point an infinite number of times at unbounded values
of $t$, causing the metric terms appearing in the ratio of radii to
effectively cancel out.

In relativity, there is no unambiguously defined global time in general.
If this is the case, the only reasonable option is to use the particle's
proper time as $t$ in Eq.~(\ref{Liap}). This approach emphasizes the
quasi-local nature of Lyapunov exponents, that is, their relation to particular
trajectories, and not (necessarily) the system as a whole. If the system has
a timelike Killing vector, and hence a global time $t$, then this may be used
in the definition of the exponents. The results of this approach correspond
to measurements of stationary observers. We choose the latter approach,
following Ref.~\cite{KV}.

In the MP case, there is an additional difficulty in that many of the
geodesics are incomplete. Particles may cross the horizon of a
black hole after finite proper time and encounter a singularity, rendering
the limit in Eq.~(\ref{Liap}) undefined. Thus, we attempt to
calculate Lyapunov exponents only for those trajectories which survive for an
infinite proper time. This is not a problem for the regions of phase space
which are nearly integrable, but for the basin boundary special techniques are
required. It is impossible to find a point which is exactly on the boundary,
and a trajectory which begins near the boundary will eventually move
away and fall into a black hole. The solution is to periodically adjust the
trajectory to ensure that it remains close to the boundary. This is achieved
by small random shifts, checking to see that the resulting trajectory
survives more than a specified proper time. We have checked that this
procedure is stable with respect to the precision used and does not result
in systematic shifts in the energy $E$ over the integration times used.
We have calculated the Lyapunov exponents for the two-black hole system using
the 4-dimensional sub-manifold of phase space, ($x$, $z$, $v_{x}$, $v_{z}$),
defined by ($L_{z}=0$, $m=\rm const$), for the initial values
$(3.33467,0.23509,0,0)$,
and found them to be $\lambda_{k}=(0.03609,0.00006,-0.00006,-0.03609)$.
The statistical uncertainty is about $0.00009$.  Note that the Lyapunov
exponents are measured in the eigendirections of the linearized evolution
operator, and do not correspond to the above coordinate system.

As in non-relativistic mechanics, the symmetries of the system impose
constraints on the Lyapunov exponents~\cite{LL}. Liouville's theorem implies
that they sum to zero; time reversal symmetry implies that they come in $+/-$
pairs; and the presence of a constant of the motion requires that one
pair of the exponents be zero. Since the phase space for our system
is eight dimensional, we require four constants of motion to ensure the
vanishing of all four pairs of Lyapunov exponents. For geometries with three
or more black holes, there will generally only be two constants of motion,
necessitating the calculation of two pairs of Lyapunov exponents.  These
results will be presented in a future study.
The two-black hole geometry furnishes three constants of motion, leaving one
pair of non-zero Lyapunov exponents as given above.

Finally, we note that recently the MP solutions have been generalized to
include a positive cosmological constant~\cite{KT},~\cite{BHKT}.
These solutions
may describe coalescing extremal black holes in a de Sitter type universe. It
would be interesting to investigate this system, and how its dynamical nature
and negative Ricci curvature might affect the chaotic structure of the
geodesics.

\end{document}